# CREASE-2D Analysis of Small Angle X-ray Scattering Data from Supramolecular Dipeptide Systems

Nitant Gupta[1#], Sri V.V.R. Akepati[2#], Simona Bianco[3], Jay Shah[1], Dave J. Adams[3*] and Arthi Jayaraman[1,2,4*]

[1] Department of Chemical and Biomolecular Engineering, University of Delaware, Newark, DE 19716, United States.

[2] Data Science Program, University of Delaware, Newark, DE 19716, United States.

[3] School of Chemistry, University of Glasgow, Glasgow, G12 8QQ, UK.

[4] Department of Materials Science and Engineering, University of Delaware, Newark, DE 19716, United States.

**# equal contributions**

* Email for correspondence: arthij@udel.edu  and dave.adams@glasgow.ac.uk

**Abstract**

In this paper, we extend a recently developed machine-learning (ML) based CREASE-2D method to analyze the entire two-dimensional (2D) scattering pattern obtained from small angle X-ray scattering measurements of supramolecular dipeptide micellar systems. Traditional analysis of such scattering data would involve use of approximate or incorrect analytical models to fit to azimuthally-averaged 1D scattering patterns that can miss the anisotropic arrangements. Analysis of the 2D scattering profiles of such micellar solutions using CREASE-2D allows us to understand both isotropic and anisotropic structural arrangements that are present in these systems of assembled dipeptides in water and in the presence of added solvents/salts. CREASE-2D outputs distributions of relevant structural features including ones that cannot be identified with existing analytical models (e.g., assembled tubes' cross-sectional eccentricity, tortuosity, orientational order). The representative three-dimensional (3D) real-space structures for the optimized values of these structural features further facilitate visualization of the structures. Through this detailed interpretation of these 2D SAXS profiles we are able to characterize the shapes of the assembled tube structures as a function of dipeptide chemistry, solution conditions with varying salts and solvents, and relative concentrations of all components. This paper demonstrates how CREASE-2D analysis of entire SAXS profiles can provide an unprecedented level of understanding of structural arrangements which has not been possible through traditional analytical model fits to the 1D SAXS data.

I. **Introduction**

Supramolecular structures formed from assembly of small molecule (low molecular weight) building blocks exhibit hierarchical structural diversity in the shapes, sizes, and pores of the assembled domains.[1-4] For example, the hierarchical structure in these supramolecular gels can include assembled tubular micelles, ribbons, rods, or fibers which interact to form complex three-dimensional (3D) network/gel structures that can immobilize the solvent(s) in the pores.[5, 6] The complexity and diversity of structures can be tailored by the choice of building block chemistry, solution conditions (pH, salts), concentration of the various components, and assembly protocol.[2] These factors impact the effective interactions between small molecules and interactions between them and components of the solution (e.g., solvent(s) and salt), in turn tuning the assembled structure at equilibrium or the kinetically trapped metastable state.[3, 6, 7]

Dipeptides are a class of small molecule building blocks (gelators) that exhibit diverse non-covalent interactions, pH sensitivity, solvent amphiphilicity, and readily assemble in solution into diverse structures. For such systems, the dipeptides assemble into micelles at high pH and then gels can be formed either directly by crosslinking of the micelles with a divalent salt or by reduction in the pH.[8-10] Understanding the micellar state prior to forming gels is critical. There have been some studies focused on this micellar state; e.g., in a recent study, the packing of the naphthalene dipeptides (2NapFF) in the micellar state was reported, revealing that the dipeptides predominantly assemble into a tubular morphology.[11] The diastereomeric (L,D)-2NapFF version was found to give nanotubes a significantly higher radius than the (L,L)-2NapFF.[11] (**Figure 1**)

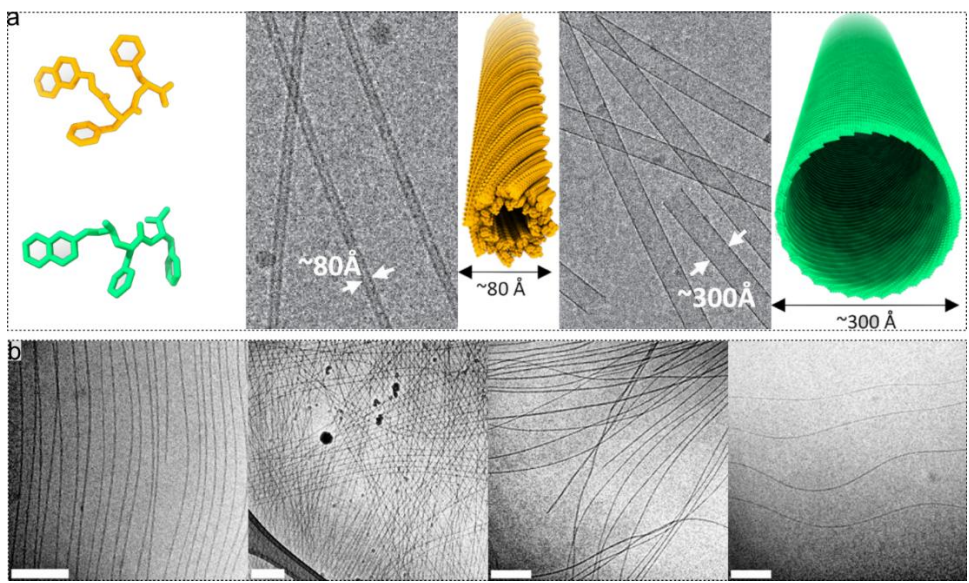

***Figure 1***: *(a) Tubular micelles resulting from differences in packing of diastereoisomers of the 2NapFF dipeptide leading to two distinct diameters: 80 Å for (L,L)-2NapFF and 300 Å for (L,D)-2NapFF. Cryo-TEM results display uniformity in diameter across their length, and amongst all tubular micelles. Reprinted from Ref.* [11] *Cell Reports Physical Science, Copyright (2024), with permission from Elsevier. (b) Cryo-TEM of micelles formed by other dipeptide after dilution and drying to display individual tubes' structural*

*features, such as bending, curving, and having orientational alignment or disorder. Reproduced from Ref.* [12] *with permission from the Royal Society of Chemistry.*

Structural characterization of the micellar structures formed by these and related dipeptides has presented some challenges. Microscopy imaging of the hydrated samples is difficult unless cryoTEM is used. Drying the sample before imaging makes it mechanically weak and likely to collapse under high vacuum environments. There also remains the question as to how much of the structure changes upon drying.[13] Even in the case where solution-state images are captured using cryoTEM, quantifying various aspects of the assembled network structure from two-dimensional images can be quite complicated (e.g., resolving the dimensions of assembled tubes and the shapes of the tubes' cross-section), and issues such as shear-alignment during blotting can easily occur.[14] In Figure 1, we present one example of tubes formed from self-assembled dipeptides. Even the variation in stereo-isomerization - from (L,L)-2NapFF to (L,D)-2NapFF – can lead to differences in the shapes and sizes of the tubular structures with the tube diameters changing from 80 to 300 Å, respectively. Images from experiments with varying solution conditions suggest that the tubular micelles are flexible with some tortuosity along the length of the tubes. Some images show tubes that are orientationally aligned while others exhibit orientationally disordered configurations. Such imaging-based information sets our expectations about possibilities of tubular morphologies assembling to form the real-space structure in their gel state. However, it remains difficult to ascertain that the same structural morphologies also exist in the bulk state, and it is extremely difficult to access such images on the gel state. Hence, we need an independent verification through an alternate characterization method, such as small angle X-ray scattering (SAXS).

Unlike microscopy, SAXS requires minimal pre-processing of the materials and captures the average three-dimensional structure of supramolecular gels and micellar solutions in their hydrated state. Data from SAXS measurements include the intensity of scattered waves for varying wave vectors I($\mathbf{q}$). In **Figures 2a-2c**, we present raw unprocessed 2D SAXS profiles for three different systems of dipeptides labeled as E1 to E3 with – a) sample E1 representing an aqueous solution of (L,D)-2NapFF (10 mg mL$^{-1}$) in presence of 10% DMSO (v/v), b) sample E2 representing a aqueous solution of 2NapIF (10 mg mL$^{-1}$) in presence of 6 molar equivalents of KCl, and c) sample E3 representing an aqueous solution of BrNapIF (10 mg mL$^{-1}$) without any additional solvents. Chemical structures of the dipeptides, necessary details of the sample preparation for these and additional samples, as well as the SAXS measurements are presented in the **Supporting Information (SI) section S.I**.

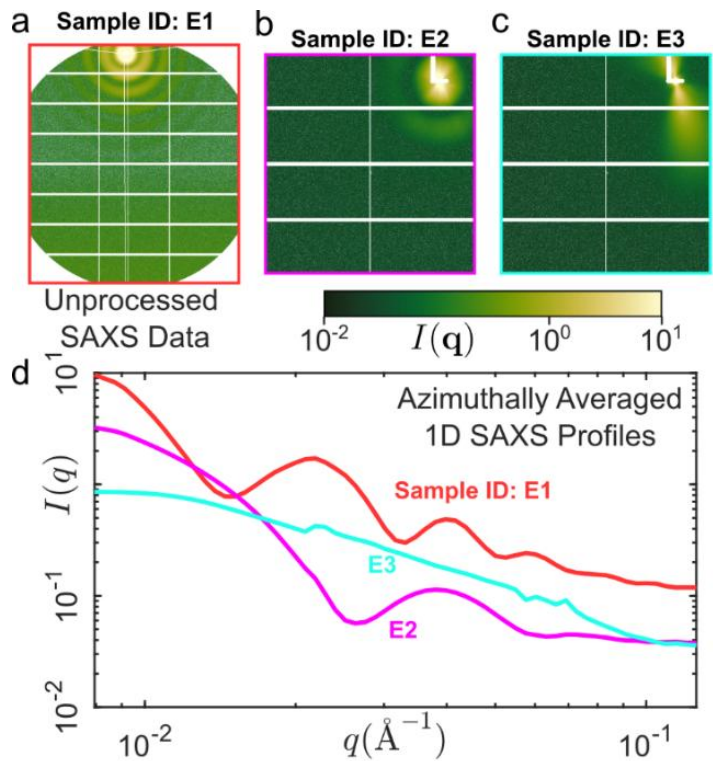

*Figure 2 (a-c) Raw unprocessed SAXS data of samples E1, E2 and E3. (d) Representative azimuthally averaged 1D SAXS profiles of samples E1-E3 comprising of various dipeptide micellar systems. Details of the conditions corresponding to these experimental samples are provided in Table S.1. in supporting information. The corresponding details of the sample preparation and SAXS measurements are presented in the supporting information section S.I.*

The I(**q**) data from SAXS holds information about the shape (or the form P(**q**)) of the tubular structures as well as the structural arrangements of the tubes [i.e., S(**q**)]. If the concentration of peptides in solution is diluted enough for tubular structures to remain isolated without further assembly or interaction, then I(**q**) ~ P(**q**). If the concentration of peptides in solution is high enough for the tubular structures to form a network or higher-order structure, then it is important to interpret the information pertaining to both P(**q**) and S(**q**) in the I(**q**)~P(**q**)S(**q**). Such multi-length scale information encoded in the scattering pattern must be interpreted to build an accurate real-space structural understanding of the tubes.

Traditional ways to interpret 2D SAXS measurements would involve plotting azimuthally averaged scattering patterns resulting in 1D scattering profiles – I(q) vs. q. In **Figure 2d**, we present the azimuthally averaged 1D profiles for the complete scattering data shown in Figures 2a – 2c. While the differences between the three systems are captured in both the 1D and 2D profiles, the anisotropic details are more prominent in the 2D profiles than 1D profiles (especially for sample E3) as these details are averaged out in the 1D profiles. This necessitates the use of computational methods like CREASE-2D [15] that interpret the entire 2D scattering pattern without any averaging along the azimuthal angles. To the best of our knowledge, there are no other methods besides

CREASE-2D that allow for interpretation of the entire 2D SAXS profile as is without any averaging along sections of or all of the azimuthal angles.

In this paper, we extend the new CREASE-2D approach to this complex system of aqueous solutions of dipeptides assembled into tube-like micelles. In the next section (**Section II**), we review the CREASE-2D method in general. In the following section (**Section III**) we describe the steps of the CREASE-2D workflow as we extend it to analyze data for the problem at hand. In **Section IV**, we present the interpreted results from the application of CREASE-2D to the SAXS data from dipeptide micellar solutions. In **Section V** we present a comparison of CREASE-2D's results with results from other approaches (simulations or traditional analytical model fits of 1D SAXS profiles) to demonstrate the power of CREASE-2D.

## II. Computational Reverse Engineering Analysis of Scattering Experiments-2D (CREASE-2D)

The CREASE method [15-18] takes as input small-angle X-ray and/or neutron scattering profiles and uses an optimization loop to evaluate and identify the values of relevant structural features whose computed scattering profile(s) matches with the input scattering profile(s). Structural features are mathematical descriptors of aspects of the structure that the user wishes to understand about their sample. CREASE-2D uses genetic algorithm (GA) for the optimization; each iteration of the GA optimization loop is referred to as a 'generation' in GA jargon. Each generation is comprised of individuals where each 'individual' has a unique set of values for the structural features that are being optimized. For each individual, we calculate its computed scattering profile using a surrogate machine learning (ML) model that has been trained on the dataset relating structural features (input to ML model) and corresponding 1D or 2D scattering pattern (output of the ML model). For every individual in a generation, we use the ML model and predict the computed scattering profile corresponding to its values of the structural features and compare it to the input scattering profile; the similarity of that individual's computed scattering and input scattering (defined as 'fitness') is quantified. Once the fitness of all individuals is calculated, genetic operations (e.g., carry-over, mutations, parents' cross-overs) are used to select and/or create high-fitness individuals in the subsequent generation's population. With increasing generations, the GA loop moves towards populations of individuals with increasing match between the computed and experimental scattering profiles (i.e., higher fitness). Once the overall fitness of the individuals in the generation converges, then the optimization is complete. The GA outputs *all* the sets of optimized structural features. These structural features are then used to create representative 3D structures to facilitate visualization of the real-space structures corresponding to the input scattering profile.

**SI Figure S1** describes this GA loop schematically. For details regarding users' choices – number of individuals in each generation, how to represent the structural features' values for each individual, decisions that dictate how many of the high fitness individuals are selected to move to the next generation versus mutated individuals – we direct the reader to previous publications of CREASE [16-18] and CREASE-2D[15].

While CREASE-2D has a similar GA loop as previous CREASE approaches, the newer aspects of CREASE 2D are its interpretation of the entire 2D scattering profiles without any azimuthal angle averaging and the inclusion of structural features that can describe the orientational alignment and/or structural anisotropy. For the system of aqueous solutions of dipeptide assembled into nanoscale tube-like micelles (the focus of this paper), we expect anisotropy to play an important role in defining the structure based on examining the microscopy images (Figure 1) and scattering profiles that also provide evidence for the presence of such anisotropy. Therefore, we choose to use CREASE-2D method to interpret this system of dipeptide tube-like micelles. CREASE-2D method will interpret key structural information but also test structural hypotheses of likelihood of certain configurations (e.g., tapes vs. cylinders, elliptical cross-section of tubes vs. circular cross-section) of dipeptide tubes. Such analyses are currently not possible with any analytical models to the best of our knowledge.

In contrast to the previous work on CREASE-2D where we presented and validated the method [15] by interpreting *in silico* 2D scattering profiles from mixtures of ellipsoidal - spherical particles, in this study we are extending CREASE-2D to real scattering data from experiments on a different system, largely comprising of dipeptide tubular viscous micellar solutions. Before we extend CREASE-2D to the SAXS data for this specific system, we need to address issues posed by the data from experimental measurements which are missing when working with pristine *in silico* 2D scattering profiles. We discuss these additional issues and the corresponding processing steps we take before analyzing this data using CREASE-2D.

In Figures 2a-2c, we see examples of raw (unprocessed) SAXS measurement data that were obtained from two different facilities as described in the SI section S.I. In these examples, the scattering intensity $I(\boldsymbol{q})$, which is recorded as a color-coded image, has some features:

(i) the position of the beam center is situated near the edge of the detector; a practice considered beneficial for isotropic specimens, to maximize the q-range for 1D profiles

(ii) the appearance of masked regions which may arise due to many reasons including beam stop shadowing, placement of slits, instrument components, etc.[19, 20]

As such, the raw data must be processed to make complete 2D scattering profiles, which is partially facilitated by the inversion symmetry of 2D scattering profiles ($I(\boldsymbol{q}) = I(-\boldsymbol{q})$). The data may further be processed by smoothing separately along the $q$ and $\theta$ directions and then replotted with $q$ in the logarithmic scale to obtain processed versions of these profiles.

**Figure 3** presents the processed 2D scattering profiles of eight experimental samples E1-E8. Details regarding the processing of the raw profiles are presented in more detail in the **SI Figure S2.** With these processing steps, the length-scale information is decoupled with higher $q$-ranges typically pertaining to the form factor P(**q**) (i.e., structural features of the individual tubes) and the lower $q$-ranges pertaining to the structure factor S(**q**) (i.e., relative spatial arrangement of tubes). It is clear that after the processing steps are completed, some of the missing information due to the masked regions may be recovered or approximated, however, a lot of the information in many other directions may remain inaccessible. When significant anisotropy is present in 2D scattering profiles, the resulting averaged 1D scattering pattern often exhibits kinks in the data due to the varying intensity values across the azimuthal angles. It follows that these datapoints make the data hard to fit *via* conventional fitting procedures, masking structural features and leading to high $\chi^2$ fitting values. One way to overcome this is to perform segment bow-tie integrations in the directions of both regions of low and high scattering intensity,[21] providing estimated 1D scattering patterns for both directions. There are some limitations in this approach as the data results from narrow ranges of azimuthal angles may not be representative of the full sample. It is also difficult to interpret if the structures observed in the low intensity regions are different populations of structures compared to the anisotropic ones. Further, as form factor fitting assumes that the structures are non-interacting and randomly oriented, this may lead to small errors in the fitting as the structures in the sample are instead aligned.

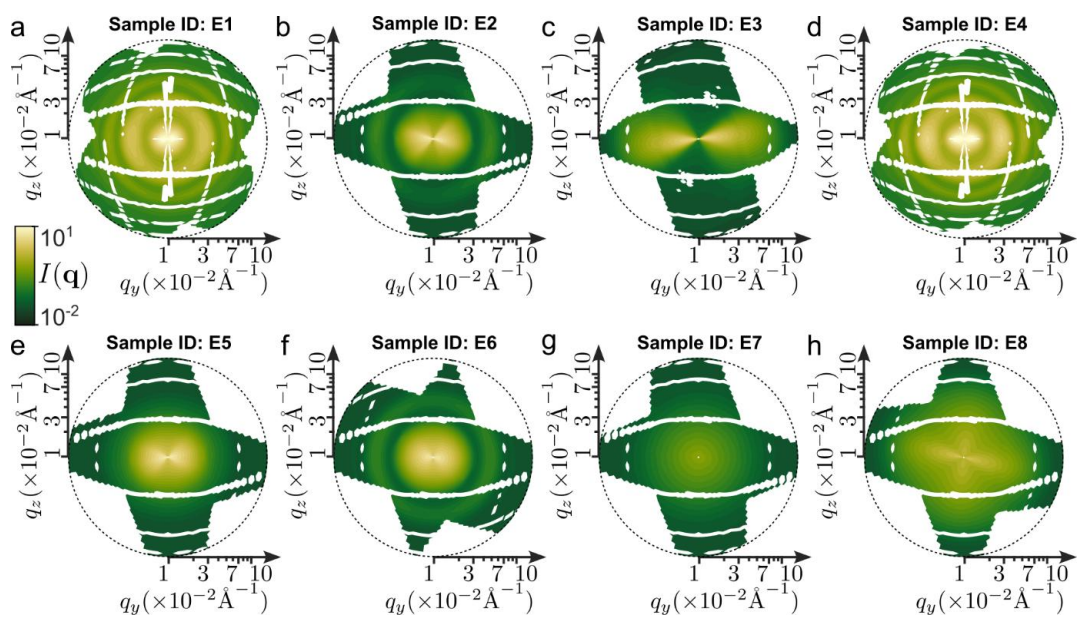

***Figure 3*** *(a-h) Processed 2D SAXS profiles for selected experimental samples, labeled as E1-E8. Details of the conditions corresponding to these eight samples are provided in Table S.1.*

In certain situations, it may be possible to repeat the measurement with slightly different position of the beam center to fill-in all the remaining information. However, often these measurements are performed at synchrotron facilities which may be conservative and strict about beamtime allocations. High operational costs, limited staff support, and limited availability of such facilities for repeated measurements can be barriers to replicating measurements. Therefore, it becomes important that the CREASE-2D method has the capability to work with the experimental data that is obtained through a single measurement per system and provides a reasonable real-space interpretation of the structure despite some missing information in the scattering pattern. We accomplish this in the GA loop by comparing only those regions of the computed scattering that have corresponding regions with available intensity information in the input SAXS profile. Provided that significant portions of the SAXS profile have available intensity information, the CREASE-2D workflow will be able to output the correct structural features that reliably represent experimental specimens.

### III. CREASE-2D Workflow

With the objective of adapting the CREASE-2D workflow to the dipeptide systems, we now outline the key steps in the interpretation of their 2D scattering profiles.

*Step 1: Identification of Structural Features for Dipeptide Micellar Solutions*

The first step is to identify the relevant structural features that are mathematical parameters that can describe relevant structural aspects (e.g., dimensions, flexibility, tortuosity, shapes) and their variations in the dipeptide micellar systems. Microscopy data as well as direct observations made by SAXS profiles can aid this effort. For example, in Figure 3 the processed 2D scattering profiles of experimental samples E1-E8 provide some guidance on the choices of length scales and presence or absence of anisotropy. The $q$ range ($10^{-2}$ to $10^{-1} Å^{-1}$) provides the important length-scale where we should direct our modeling effort, which in this case is $\sim 60 - 600$ Å. Mainly, we see the occurrence of a single peak versus many peaks in Figure 2d, which is indicated by one ring versus many rings around the center spot in Figure 3, correlated to the different diameters of the dipeptide tubes. We also notice that some of the profiles, especially sample E3, but also samples E2, E5, and E8, have high anisotropy indicated by the rotational asymmetry in their 2D profiles.

We use nine structural features to describe our dipeptide systems. A brief description of these structural features and their expected relationship to realistic structural aspects are presented in **Table 1**. We describe our rationale for choosing each of the structural features and how it relates to relevant physical aspects of the structure.

**Table 1**. *Chosen structural features to represent the dipeptide assembled tubes, their symbols, definition, and what aspect of the structure they capture*

| Parameter description | Defined Structural Feature | Symbol |
|---|---|---|
| Parameters that describe the shape and size of individual dipeptides' assembled tubes. | Circular tube diameter | $D$ |
| | Mean cross-sectional eccentricity | $e_\mu$ |
| | Normalized standard deviation of cross-sectional eccentricity | $e_\sigma$ |
| Parameters that describe the orientational order and alignment of tubes | Mean orientation angle | $\omega$ |
| | Orientational anisotropy parameter | $\kappa$ |
| Parameters that describe the tortuosity and stiffness of individual tubes by influencing the tube contour. | Herd cone angle | $\alpha$ |
| | Diameter of herding tube | $d_h$ |
| | Length of herding tube | $l_h$ |
| | Number of extra nodes inside a herd | $n_x$ |

The microscopy images in Figure 1 from our past work [11] suggest that the assembled tubes have uniform diameters near the range of $80 - 300$ Å; so, we consider that tubes must be represented with a **uniform tube diameter $D$ (first structural feature**) and neglect dispersity in the diameter based on this evidence. We also ignore the use of shell thickness as a parameter because we did not find a significant effect of that parameter during the calculation of scattering profiles. However, in the 2D scattering profiles in Figure 3, we observe broadening of some peaks in the scattering profiles which suggests the possibility of dispersity in this system, likely in some other structural feature. As the microscopy images show uniform diameters, we hypothesize that the dispersity is likely in the tubes' cross-sectional shape. For this reason, we introduce the concept of tube cross-sectional eccentricity $e$ which varies between 0 (perfect circle) and 1 (flat ribbon) as also shown in **Figure 4a.** Additionally, we expect that the cross-sectional eccentricity can also vary along the length of the tube, assuming that the circumference of the tube is conserved. For this reason, we define $D$ as the circular diameter of the tube, such that its circumference $\pi D$ is always conserved at any value of eccentricity $e$. We then use a $\beta$-distribution to model the variation of eccentricity with a mean $e_\mu$ and a normalized standard deviation $e_\sigma$ (for numerical convenience). $e_\sigma$ always lies between 0 indicating no dispersity, and 1 indicating maximum dispersity (with the maximum value dependent on $e_\mu$ as is typical in a $\beta$-distribution). Therefore, the **second and third structural features are $e_\mu$ and $e_\sigma$.** The variation in the distribution of eccentricity for four configurations are showcased along with the histograms of their distribution and the representative snapshots in **Figure 4b.** The similarity in the snapshots of **Figure 4c** is meant to showcase that these parameters can all be independently varied, such that while tube placements, diameters and contours are similar, only the cross-sectional eccentricity varies as seen in **Figure 4d**, which is an important consideration for its effect on the calculated scattering profile.

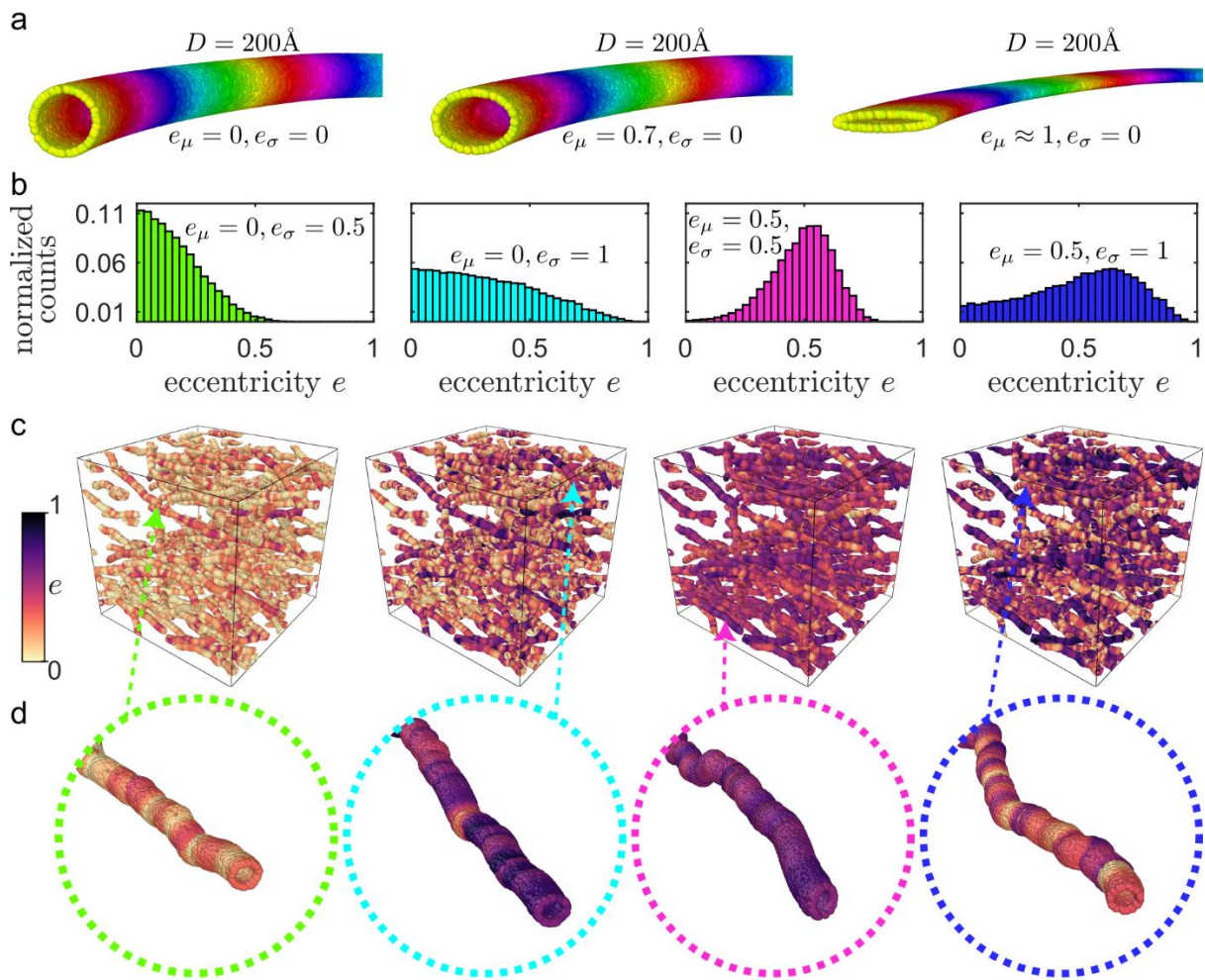

*Figure 4*: *(a) Representative snapshots of single tubes with same diameter but varying cross-section shape (i.e., varying mean eccentricity $e_\mu$ and zero dispersity $e_\sigma = 0$). (b-d) Representative snapshots of systems with varying mean and standard deviation in eccentricity - $e_\mu$ and $e_\sigma$, with (b) showing histograms of eccentricity, (c) complete snapshot of the dipeptide system and (d) close-up of a single tube from the system. The color-bar on the left codes the varying cross-sectional eccentricity along the length as well as between different tubes in (c) and (d).*

As has been demonstrated in our earlier work [15, 22], the anisotropy observed in the 2D profiles, can be expressed by the quantification of orientational order, which is facilitated by using the von Mises Fisher (vMF) distribution. The vMF distribution models the extent of orientational alignment of all dipeptide tubes giving us **two more structural features, identified as $\omega$ and $\kappa$**, which respectively correspond to the mean orientational angle of the tubes and the orientational anisotropy parameter, as defined in our previous work [15, 22]. We can associate high $\kappa$ with high orientational alignment and low $\kappa$ with orientationally isotropic arrangement of tubes.

However, the idea of orientation of the tube may be somewhat convoluted, especially when the tubes are highly flexible and can be curled or bent into various conformations. To address this, we must first establish a mathematically consistent method to model the tube's contours and define precise parameters that influence the variations in tube conformations. These parameters may be closely related to tube tortuosity and stiffness. Thus, before we identify the remaining structural features, we will describe how to create tubular geometries in three-dimensional space corresponding to the structural variations of the dipeptide tubes.

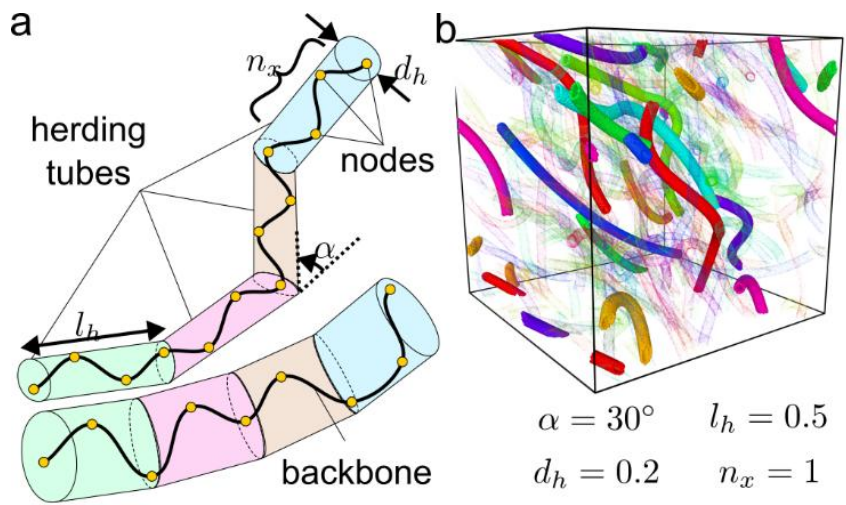

*Figure 5*: (a) Schematics demonstrating the role of 4 structural features ($\alpha, d_h, l_h$ and $n_x$) related to herds and nodes. (b) Snapshot of a reference system with fixed values of the 4 structural features. Many of the tubes in the system are made translucent to clearly visualize the effect of the 4 features.

We model the tube contour, referred to as the backbone representing the tube axis, using smooth spatial curves that are interpolated over a sequence of points, that we call nodes as shown in **Figure 5a**. We use 3D spline interpolation over these nodes using the scheme presented by Yuksel et. al.[23]. Once the tube axis is sufficiently modeled by the backbone, we then use a series of cylindrical particles, with elliptical cross-section, to model the shape of the dipeptide tubes. These particles facilitate the visualization of the final geometries of the tube morphology.

The placement of nodes is strategically selected to ensure that structural aspects like orientation of the tubes as well as their tortuosity and stiffness can be reliably modeled. This is accomplished by first envisioning a connected set of long rigid cylinders, which we refer to as 'herding tubes' or 'herds', as shown in Figure 5a, which shall limit the spatial extent of all the nodes inside them (the reader can imagine this process to be like sheep herded by fences). Inside each of these herds, nodes can be placed randomly along the diameter of the herds, but sequentially along the length of the herd; therefore, the dimensions (diameter and length) of the herds, as well as the angle between connected herding tubes become important parameters that control the stiffness and tortuosity of the modeled dipeptide tubes. We choose **four structural features** that correspond to the "herds" and nodes, used to influence the tortuosity and stiffness of the dipeptide tubes. These are, namely, the **herd cone angle $\alpha$**, the **diameter of herding tubes $d_h$**, **length of the herding tubes $l_h$**, and the **number of extra nodes in a herding tube $n_x$**. These structural features are all depicted in the schematic in Figure 5a.

The herd cone angle $\alpha$, which varies between 0° and 90°, denotes the spherical cone at the end of a herding tube within which the orientation of the next connecting herding tube randomly resides. When, $\alpha = 0°$, the spherical cone is a straight line pointing along the same direction as the previous herding tube, such that the resulting arrangement will show all connected herding tubes pointing in one direction. However, when $\alpha = 90°$, the spherical cone is now a hemisphere, and the next herding tube can essentially point in any direction that is away from the previous herding tube ($\alpha > 90°$ is not permitted). The diameter $d_h$ and length $l_h$ of the herding tube can directly control the curviness of the spatial curve by tightening or relaxing the random placement of the nodes within the herding tubes. The number of extra nodes $n_x$ in herding tubes refers to how many nodes are placed within each herding tube. By default, at the connection of two herding tubes a node is always present (for the first and the last herding tube, the ends which have no connection also have a node each). Therefore, $n_x$ controls how many more nodes are included within a herding tube, between those already present at the ends of each herding tube. Thus $n_x$ can be 0, indicating no extra nodes are present, and can count to any number as long as the configuration remains reasonably meaningful. In this work, we restrict $n_x \leq 5$.

In **Figure 5b**, a representative system of dipeptide tubes is shown at chosen values for $\alpha$, $d_h$, $l_h$ and $n_x$, and in **SI Figure S3** we show the effect of varying these four structural features as compared to this reference system. We can see that each of these structural features can individually influence the stiffness and the tortuosity of the tubes. For example, lower values of $\alpha$ mean that the assembled tubes are extended than folded; higher values of $l_h$ suggest larger persistence lengths or stiffness. Higher values of $d_h$ accompanied with higher values of $\alpha$ suggest higher tortuosity. However, it is difficult to decouple their individual effects, or to define other better descriptors for achieving the same goal. We will rely on the CREASE-2D GA algorithm to optimize these structural features to correctly match the scattering profiles and will always have the option to generate the 3D structure and visualize how these four values impact the tube morphologies.

Having described the framework for implementing the tube geometries, we can now quantify the orientations of the tubes as from the orientation of the first herd of each dipeptide tube, which then influences the overall orientation of the dipeptide tubes. The mean orientational angle $\omega$ is the angle between the mean direction of tube orientation in the $xz$ plane, measured with respect to the $z$-axis. In **SI Figure S4**, we have demonstrated how the variation of $\omega$ at different $\kappa$ values change the structure of the dipeptide tubes. It is important to note that role of the $\kappa$ parameter is to introduce orientational anisotropy *between* different dipeptide tubes and *not within* the contours of each tube. Thus, only when $\alpha$ is close to 0°, high $\kappa$ will mean that all tubes are straight and point along the same direction. For low values of $\kappa$, the system becomes orientationally isotropic, and $\omega$ loses its meaning. It remains entirely possible that for configurations where the tubes may bend or curve excessively, the quantification of orientational order is less meaningful, and for such cases, the values of $\omega$ and $\kappa$ may be neglected, as the variations of other structural features may entirely dominate their computed 2D scattering profiles.

Lastly, we address one specific issue regarding mean orientation of anisotropic systems of dipeptide tubes. So far, in our identified list of structural features, we have only used one angle $\omega$ to describe the mean orientation of the tubes. However, the mean orientation should be a 3D vector and should more generally be described by two angles pertaining to the spherical coordinate system (where the radial coordinate is ignored). We demonstrate in **SI Figure S5** how one of these two angles can be eliminated, by looking at scattering profiles from the $yz$ and the $xz$ planes for the same sets of structures while varying $\omega$. For this demonstration, we purposefully chose a system of relatively straight tubes. When $\omega = 0°$, the scattering profiles in the $yz$ and the $xz$ planes look identical, because from the 2D snapshot, the two configurations look identical. For all other values of $\omega$, the 2D snapshots in the $yz$ and the $xz$ planes look different, and therefore the scattering profiles also differ. A keen observer would notice that since $\omega$ lies in the $xz$, the rotation of the mean orientation of the tubes, only causes the $xz$ scattering profiles to rotate as well, while for $yz$ scattering profiles, the rotations are out-of-plane, and therefore the scattering profile appears to become less anisotropic as $\omega$ increases. Therefore, we can conclude from this demonstration that any in-plane rotation (with respect to the scattering calculations) of the mean orientation of the tubes can also be achieved by simply rotating the calculated scattering profile, and only the out-of-plane rotations are important to model as they directly change the scattering profiles. For this reason, in the processing of experimental data, we also identify the direction of anisotropy by directly observing the experimental scattering profile and then rotating them to align with the $q_y$ axis, thereby eliminating the need to generate scattering dataset for all other in-plane mean orientations of the tubes.

### *Step 2: Generation of 3D structures for Varying Values of the Structural Features*

Having identified all the 9 relevant structural features (**Table 1**), we employ a novel computational method, heavily adapted from the CASGAP method[22], to generate the 3D structures of the dipeptide tubes for various values of the structural features. The purpose of doing this structure creation is to compute corresponding scattering profiles and generate data to train an ML model to link structural features → scattering profile.

### *Step 3: Calculating 2D Scattering Profiles for the 3D Structures Generated*

As described previously, the 3D structures of the dipeptide tubes are generated as a sequence of cylindrical particles with elliptical cross-sections. However, to calculate the scattering profiles from these geometries, the tubes need to have a hollow core. We achieve this by placing random point scatterers in the shells of the cylindrical particles, ensuring that geometry of the tube is replicated with hollow cores. The 2D scattering profiles are then calculated by using a modified form of the Debye scattering equation, where we first calculate the complex scattering amplitude using parallel processing and then squaring the final combined scattering amplitude. For increased computational efficiency in terms of speed and memory, the scattering calculations are further parallelized by splitting the entire structure into smaller chunks, as shown in **Figure 6a**, which are obtained by distributing all the tubes into only one of the chunks before placing point scatterers.

In this way, each chunk contains only some of the tubes from the entire structure, enabling us to use a high density of point scatterers to generate good quality (low noise) computations of scattering amplitude $A_i(\boldsymbol{q})$. Ultimately, the individual contributions from each chunk of the scattering amplitude $A_i(\boldsymbol{q})$ is added together and then squared to obtain the complete scattering profile $I(\boldsymbol{q}) = |\sum_i A_i(q)|^2$. After computation, the individual chunks can be deleted to free the memory requirements for subsequent computations.

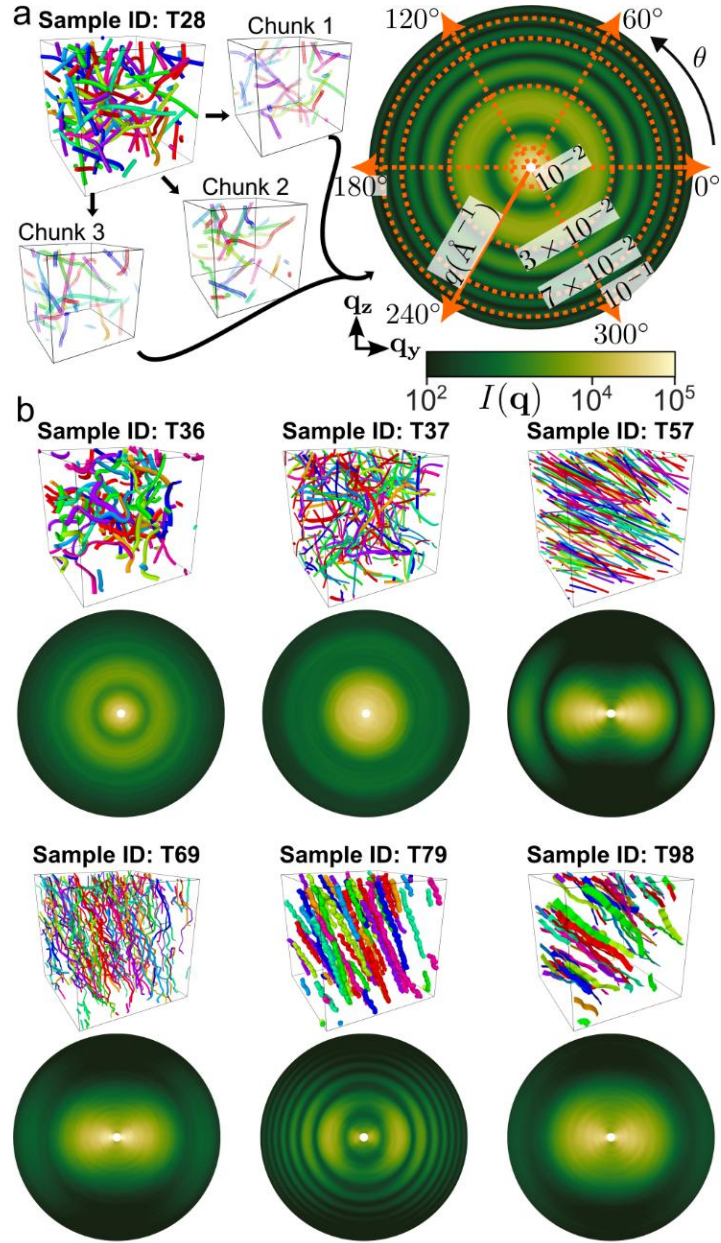

*Figure 6*: *(a) Snapshot of sample T28 (with particles) and converted into three chunks of slightly more than 80,000 scatterers each. Contributions from each chunk are then combined to evaluate the 2D scattering profile (on the right). The axis labels are provided for completeness (and omitted from all other profiles to improve visibility of the patterns in the 2D scattering profiles). (b) Snapshots and 2D scattering*

*profiles of representative samples from the training dataset: T36, T37, T57, T69, T79, and T98. A complete list of structural features for these samples is provided in Table 2.*

We found that scattering profile from only one representative structure can be noisy and can heavily influence the next step of training of the machine learning model. We therefore reduce the noise in the scattering profile by first averaging the calculated scattering profiles over five replicates of the system keeping all nine structural features the same. We then further smoothen the scattering profiles in the $q$ and $\theta$ directions to obtain the 2D scattering plot shown in Figure 6a. Few other examples of the computed scattering profiles along with their representative 3D structures are also shown in **Figure 6b** and a complete list of their structural features' values are provided in **Table 2**.

**Table 2**: *List of structural features for selected in-silico samples used for ML model training (Figure 4) and validation (Figure 6).*

| Sample ID | $D$ | $e_\mu$ | $e_\sigma$ | $\omega$ (degrees) | $\kappa$ (log scale) | $\alpha$ (degrees) | $d_h$ | $l_h$ | $n_x$ |
|---|---|---|---|---|---|---|---|---|---|
| T28 | 340 | 0 | 0.04 | 5 | -0.07 | 65 | 0.08 | 0.6 | 2 |
| T36 | 340 | 0.8 | 0.48 | 45 | -2.28 | 55 | 0.48 | 0.24 | 1 |
| T37 | 180 | 0.88 | 0.52 | 20 | -1.79 | 80 | 0.16 | 0.56 | 2 |
| T57 | 120 | 0.84 | 0.16 | 55 | 2.84 | 10 | 0.08 | 0.96 | 0 |
| T69 | 110 | 0.92 | 0 | 15 | 2.18 | 35 | 0.16 | 0.08 | 1 |
| T79 | 380 | 0.24 | 0.08 | 25 | 2.37 | 5 | 0.48 | 0.12 | 4 |
| T98 | 340 | 1 | 0.72 | 45 | 3.91 | 15 | 0.12 | 0.24 | 4 |
| V10 | 280 | 0.92 | 0.92 | 30 | -1.72 | 35 | 0.4 | 0.44 | 2 |
| V177 | 120 | 0.68 | 0.56 | 10 | 1.23 | 25 | 0.24 | 0.52 | 0 |
| V185 | 120 | 0.24 | 0.48 | 30 | 2.42 | 10 | 0.4 | 0.04 | 0 |
| V280 | 300 | 0.24 | 0.48 | 65 | -0.41 | 70 | 0.32 | 0.44 | 2 |

In experimental measurements, we expect that the spatial arrangements of the tubes can also affect the scattering profile, and we will need to be account for that larger length-scale structural arrangement in the interpretation of the SAXS profiles. While it is possible to construct a model system with tubes packed into different possible assembled structures and their corresponding structural features[17], here we choose to adopt an analytical model known for networks [e.g., Debye-Anderson-Brumberger (DAB) model[24, 25]] to incorporate such larger length-scale contributions to the scattering profile, when applying CREASE-2D. Specifically, we use the correlation length $\zeta$, in DAB models to capture quantitatively the associated larger length scale structural feature (e.g., pore sizes in networks) that is present the SAXS measurement.

### *Step 4: Machine Learning Model Linking Structural Features to Scattering Profiles*

By completing Steps 2 and 3, we generate a large dataset of structural features and corresponding scattering profiles, which can be used to train a surrogate machine learning (ML) model. The ML model predicts computed scattering profiles for input values of the nine structural features describing dipeptide assembled tubes. Our past studies using CREASE and CREASE-2D have shown that the use of such a surrogate ML model for calculating scattering profiles accelerates the GA optimization loop within the CREASE-2D framework by orders of magnitude over the Debye-calculation using 3D structures.

In our previous work [15] we found that XGBoost [26] based ML model performed incredibly well in predicting the anisotropic scattering profiles for the system of ellipsoidal particles. Leveraging that knowledge, we train an XGBoost model using portion of the generated dataset of 5000 samples [where each sample is a set of values for the nine structural features and its corresponding computed scattering profile I(q, azimuthal angle)]. We split the dataset into 4000 and 1000 samples for model training and validation, respectively. Moving forward, any sample that is used for ML model training is labeled starting with T and any sample used for ML model validation is labeled with a V. Keeping with this nomenclature, we have labeled the experimental samples with E.

Before we use the ML model in the CREASE-2D GA loop, we describe aspects of the ML model including the learning curve, the 'importance' histogram of the structural features, and its performance. In **Figure 7a**, we display the Pearson correlation matrix, which denotes the correlation observed between all the inputs that will be assigned to the XGBoost model. This includes all the nine structural features, as well as the $(q, \theta)$ values at which $I(q, \theta)$ will be evaluated. The last column of the correlation matrix also shows the correlation of all the inputs with the expected output $I(q, \theta)$. We find that all correlations remain below 0.1 in magnitude, indicating weak correlation, except for the correlation between $q$ and $I(q, \theta)$ which are strongly anti-correlated with a value of -0.88. This is visually apparent after a quick look at any of the scattering profiles, where low $q$ always has high intensity. Having a low correlation between the inputs is considered a desirable attribute of the training data, as the parameters will have relatively independent influence on the ML predicted values.

In **Figure 7b**, we showcase how changing the size of the training dataset influences the learning score of the ML model. We initially started with a training dataset of 2000 samples and found that after completely training on the dataset the $R^2$ score evaluated on the validation dataset saturated to 0.96. Upon increasing the training dataset to 3000 samples, a small improvement in performance was observed over the same validation dataset. However, with the final training dataset of 4000 samples, a marked improvement to $R^2$ score ~0.98 is observed. It is possible that with further increase in the dataset size, a more significant improvement may be achieved, however, we consider this performance sufficient for our current purpose. For the XGBoost model trained on the 4000 samples, we show the feature importance in **Figure 7c**, which represents how the ML model interprets the input data, and correctly attributes $q$ with the highest importance, while all the other parameters are found to have evenly distributed importance.

In **Figure 7d**, the performance of the ML model trained on 4000 samples is tested on 1000 samples of validation data and plotted the structural similarity index measure (SSIM) scores. SSIM evaluates the similarity between two image data, and assigns a value between -1 to 1, where -1 indicates very dissimilar, 0 indicating no correlation and 1 indicating that the two images are identical. The SSIM scores for the ML model are obtained by directly comparing the original scattering profiles from the validation dataset to the predicted scattering profile. Some of the representative examples of original and predicted scattering profiles are shown in **Figure 7e**, and their corresponding structural features are tabulated in Table 2.

For the samples with a high SSIM score, the visual similarity between the original and predicted scattering profiles can be observed. For low SSIM score samples, there may be artifacts in the predicted scattering profile that diminish the visual similarities between the original and predicted profiles. It is possible that by increasing the training set and including more examples of samples with poorer ML performance the SSIM scores can be further improved for all possible values of structural features.

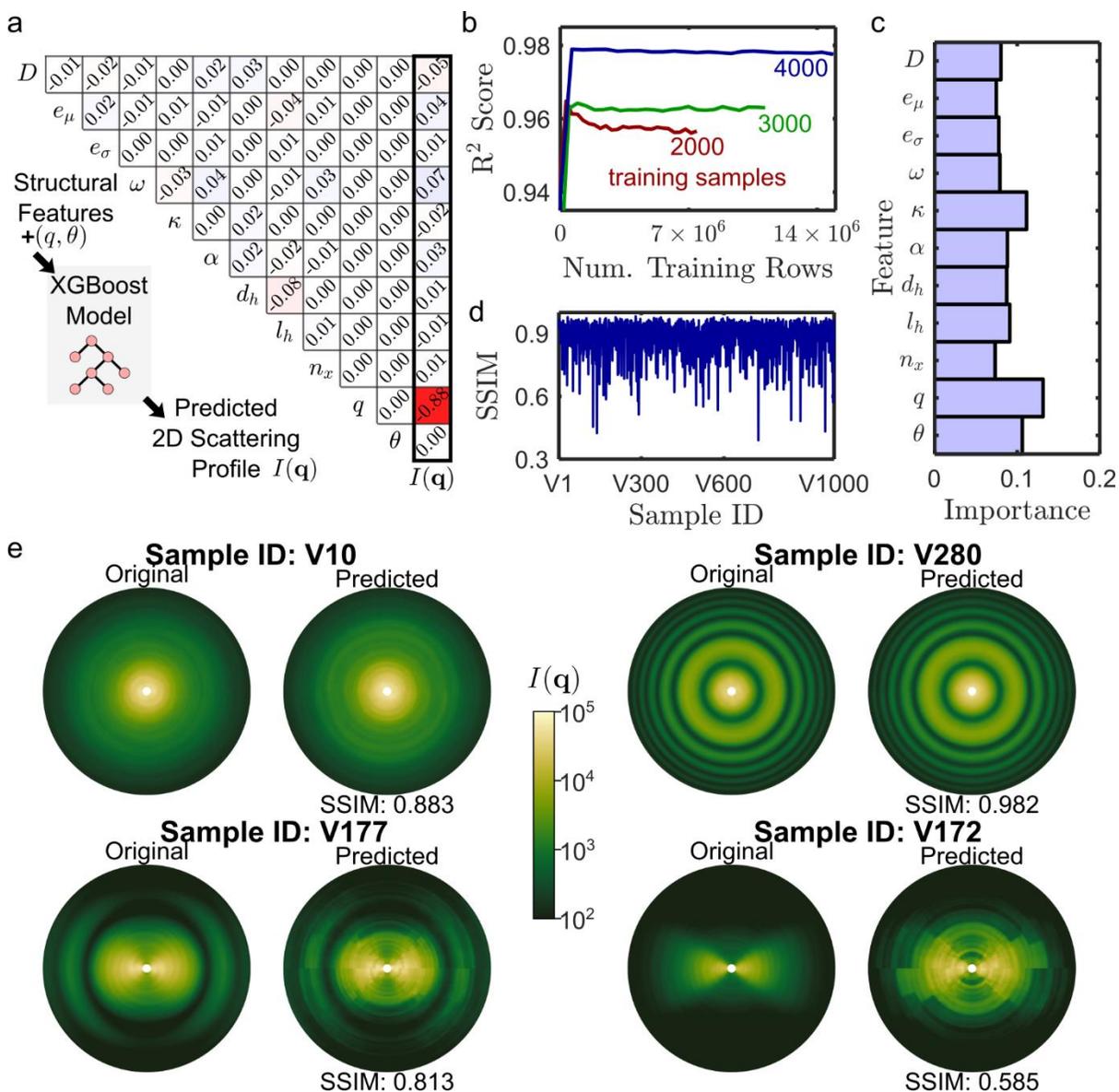

***Figure 7***: *(a) Correlation matrix for the inputs to the XGBoost ML model comprising of the 9 structural features along with q and θ. The last column indicates the correlation with the output I(**q**). (b) Learning curve during training of the XGBoost model where the size of the training dataset was varied between 2000, 3000 and 4000 samples. (c) Importance histogram for each input feature to the XGBoost model trained with 4000 samples. (d) Performance of the XGBoost model using the structural similarity index measure (SSIM) scores for all 1000 validation samples. (e) Original and predicted scattering profiles for a few samples (specified in Table 1) from the validation dataset with their SSIM scores indicating the quality of their match.*

*Step 5: Performance of the Genetic Algorithm (GA) Loop with ML Model*

The final step in CREASE-2D implementation is to incorporate the ML model inside the GA loop as described in previous CREASE publications [15, 16, 18]. We follow a similar protocol as our previous CREASE-2D GA implementation [15], using a continuous parameter GA, normalizing the structural features to genes which vary between 0 and 1, and use adaptive mutation. In this work, we use a population size of 300 individuals, with 100 parents and a total of 300 generations.

To demonstrate how we evaluate the performance of the CREASE-2D GA optimization, we share two specific samples from the *in-silico* validation dataset (of 1000 samples), namely sample V280 and V172. The V280 sample has a relatively high ML performance with an SSIM score of 0.982 (as seen in Figure 7d). On the other hand, the V172 sample has a particularly low ML performance with an SSIM score of 0.585. We describe the inner workings of the GA loop and the multiple GA runs' outputs for these two contrasting cases next.

For the V280 sample that has a high ML performance, (**Figure 8**) we show the original scattering profiles along with their 3D structures (Figure 8a) and show the evolution of the best fitness for twenty-five independent GA runs (Figure 8b); we compare the outputs, i.e., the best fit individual, from twenty-five independent GA runs using a dendrogram plot (e.g., Figure 8c). The dendrogram plot is a hierarchical clustering [27] diagram providing a convenient representation comparing the similarity (Euclidean distance) between the optimized (best) individuals (i.e., sets of values of the structural features) from the 25 GA run; the dendrogram allows us to see which individuals are similar and which are dissimilar. We also show the output from three GA runs' outputs (out of the twenty-five GA runs); two of these outputs are the most distant from each other and the third closely matches the original set of structural features (the input). It is clear from the presented information that for V280, where the ML performance is high, the fitness evolutions are similar, and the three outputs essentially represent similar optimized values for the structural features (see table in Figure 8d).

For the V172 sample that has a low ML performance, (**SI Figure S6**), the fitness evolution for all twenty-five runs is distinct, and the three outputs (the table of values of structural features) exhibit dissimilarities. For example, the "original" *in-silico* structure's diameter is 170 Å, while the three CREASE-2D GA runs output three different values of the optimized diameters – 111 Å, 216 Å, and 391 Å. Curiously, the fitness (SSIM) scores for all GA runs for sample V172 is above 0.8, which is higher than the ML performance value i.e., 0.585.

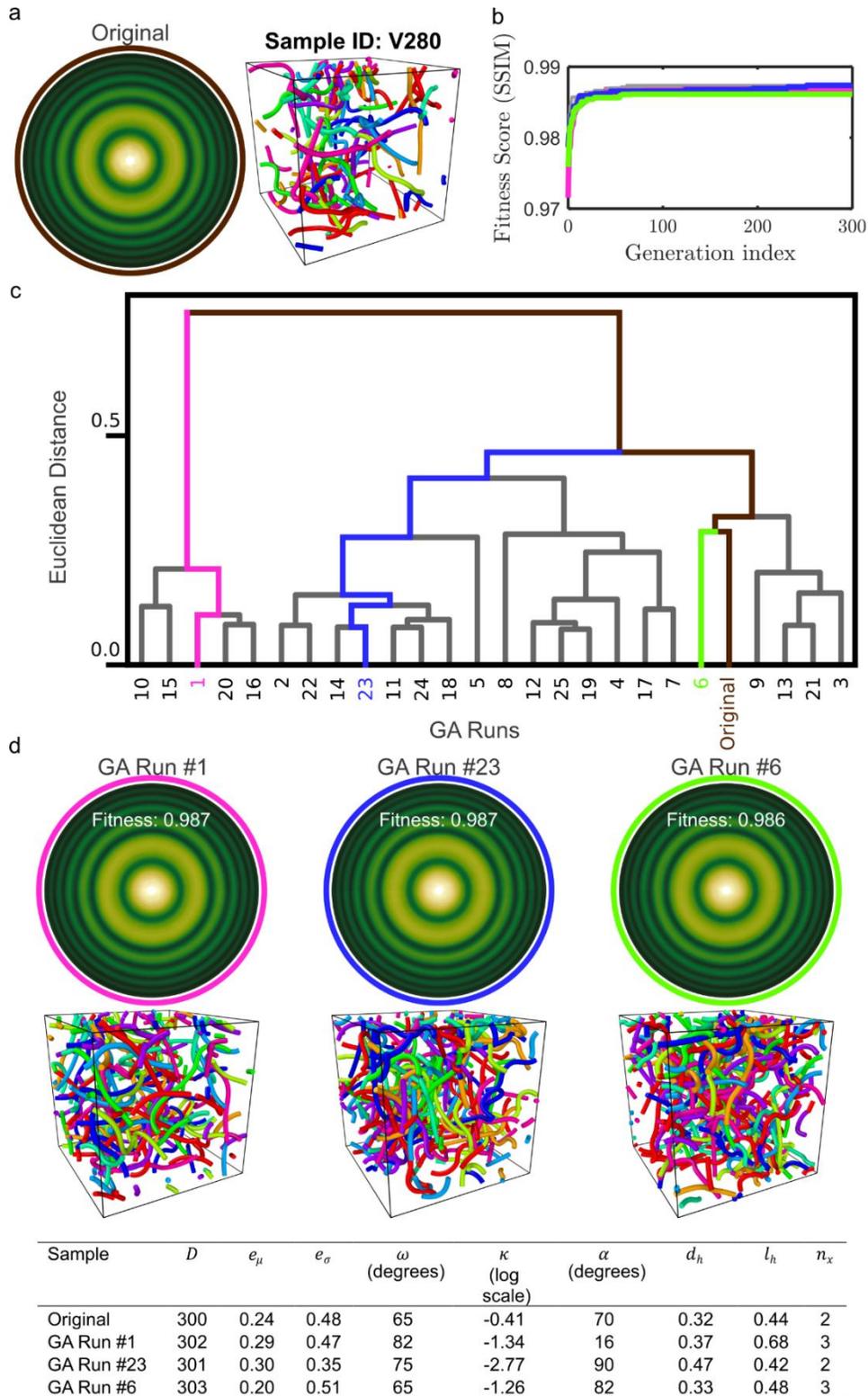

**Figure 8**: **CREASE-2D performance for the Sample V280.** *(a) Original 2D scattering profile and structure of Sample V280. (b) Evolution of best fitness during 25 independent GA runs. (c) Dendrogram*

*plot of the 25 GA runs showing similarities and differences between the outputs of the 25 GA runs. (d) CREASE-2D output of GA Run#1, #23 and #6, with best fit scattering profiles, reconstructed 3D structures and all their structural features. Among these chosen GA runs, #1 (magenta) and #23 (blue) are the most different based on the Euclidean distance between their predicted structural features. GA run #6 (green) is the closest match to the original. Color coding is used in (b), (c) and (d) to associate the plots with respect to 3 chosen GA runs in (d).*

## IV. CREASE-2D Workflow Applied to the SAXS Data

Having validated the CREASE-2D performance on *in-silico* data, we are in a position to apply this method on the 2D scattering profiles for all experimental samples E1-E8 whose SAXS profiles are shown in Figure 3 and sample details are in **SI Table S.1**. Before we discuss the results from CREASE-2D, we share some additional considerations that are necessary for the CREASE-2D GA optimization to work successfully with the experimental data. These considerations are stated below:

(1) As we noted in the beginning, all experimental data need to be processed and may still be heavily obscured by masked regions which lead to missing information in the 2D scattering profiles. In contrast, for the computed scattering profiles, all $q$ and $\theta$ values remain accessible. Therefore, during the GA optimization to make a valid comparison between the computed scattering and input SAXS profile, the mask from the experimental scattering profiles must be applied to the computed scattering profile before evaluating their fitness during the GA optimization.

(2) During the processing of the SAXS 2D scattering profiles, it is important to identify the direction of anisotropy (which is usually done by identifying the azimuthal angle at which overall scattering intensity peaks). After noting this azimuthal angle, we rotate the entire 2D scattering profile, so that the direction of anisotropy aligns with the $y$-axis of $yz$ frame (as can be seen in Figure 3), as this enables us to use only one structural feature $\omega$ to quantify the mean orientation of the dipeptide assemblies.

(3) Keen readers may have noticed that the scattering intensity of the 2D profiles for experimental samples lay in the range of $10^{-2}$ to $10^1$ (as can be seen from the color bars in Figure 2 and Figure 3). However, for the *in-silico* samples (as seen from the color bars in Figure 6 and Figure 7), the scattering intensity lies in the range of $10^2$ to $10^5$. This is due to several reasons, such as unassigned scattering length densities of constituent chemistries, incompatible volume fraction of the tubes between experiment and in silico structures, presence of background scattering in experimental data, and the arbitrary number of point scatterers used to model the structure of the tubes. Although, all of these factors may be independently controlled to bridge the mismatch in intensities, we do not consider those factors to be important in understanding the nine structural features for the experimental samples, therefore those factors will not be useful in reconstructing the 3D representation. Instead, we use two constants, $a$ and $b$, such that $a$ is a scaling factor and $b$ is a positive shift that enables the quantitative comparison of intensity values of SAXS and the computed scattering profiles during the GA optimization loop and evolve the GA individuals towards high fitness (i.e., good match).

(4) Finally, even after adequate scaling and shifting of the computed scattering profiles, to make the two scattering intensities comparable, we found that the two scattering profiles may still not match well during the GA optimization loop. This is because we need to account for the scattering contribution of spatial arrangements of the dipeptide tubes which we had thus far neglected in our *in-silico* data. As we discussed in Step 3, we employ the use of the DAB network model which enables us to use two more parameters, $S$ (scaling factor) and $\zeta$ (correlation length), such that we can add a correction term to the computed scattering profile after scaling and shifting the scattering profile. The correction term is given as:

$$I_{\text{correction}}(q,\theta) = S \frac{\zeta^3}{(1+q\zeta^2)^2}$$

When all the four considerations listed above are incorporated in the GA optimization look, the computed scattering profile $I_{comp}$ that is obtained from the ML model is modified before comparison with the SAXS profile; we now define this modified computed profile as $I_{comp\_new}$, and its complete expression is given as:

$$I_{comp\_new}(q,\theta) = aI_{comp}(q,\theta) + b + I_{\text{correction}}(q,\theta)$$

Consequently, during the CREASE-2D GA optimization loop, along with the 9 structural features, we also use the four additional parameters: $a$, $b$, $S$ and $\zeta$, to obtain a total of thirteen genes, that are optimized.

For each 2D SAXS profile from samples E1 – E8, we run twenty-five independent CREASE-2D GA runs. We present the optimized 'best' individual's values of structural features for all twenty-five GA runs, in **SI Tables S.3. – S.10**. These tables show the extent of degeneracy in the optimized values of structural features; in some cases, we see significant variability in the results from the twenty-five GA runs and in some cases the results converge to a few prominent structural interpretations. To visualize the degeneracy or clustering of the optimized structural interpretations from these twenty-five GA runs, we present dendrograms for all 8 samples **in SI Figure S7**. These dendrograms show in similar colors the clusters of similar structural interpretations.

Using this information in SI Figure S7 and the corresponding quantitative information in Tables S.3 – S.10, we describe our understanding of how varying dipeptide, salt type and concentration, and solvent(s) affects the assembled hierarchical structures. When we have multiple possible structural interpretations that vary significantly, then we compare some of the output structural features from CREASE-2D with information from other characterization data (e.g., relevant CryoTEM images like those shown in Figure 1 and analysis of the 1D SAXS patterns (**SI section S.VI.**)

First, we share some general trends obtained from CREASE-2D analysis. We notice that samples E2, E3, E5, E6, and E8 have a high value of orientational anisotropy with $\kappa > 10^2$; readers should note the positive values of $\log(\kappa)$ for these samples in SI Tables S.3 – S.10). This indicates that all these systems exhibit structural anisotropy with varying degrees of orientational alignment. Samples E3, E5 and E8, all have a low value of $\alpha$ which means their real-space structures have well-aligned nearly straight tubes. **In SI Figure S8**, we can observe some of these aspects in one representative 3D structure corresponding to values from one of the GA runs for each of these samples. Next, to understand the science of how varying dipeptide, salt type and concentration, and solvent(s) affect the assembled hierarchical structures, we compare specific sets of samples.

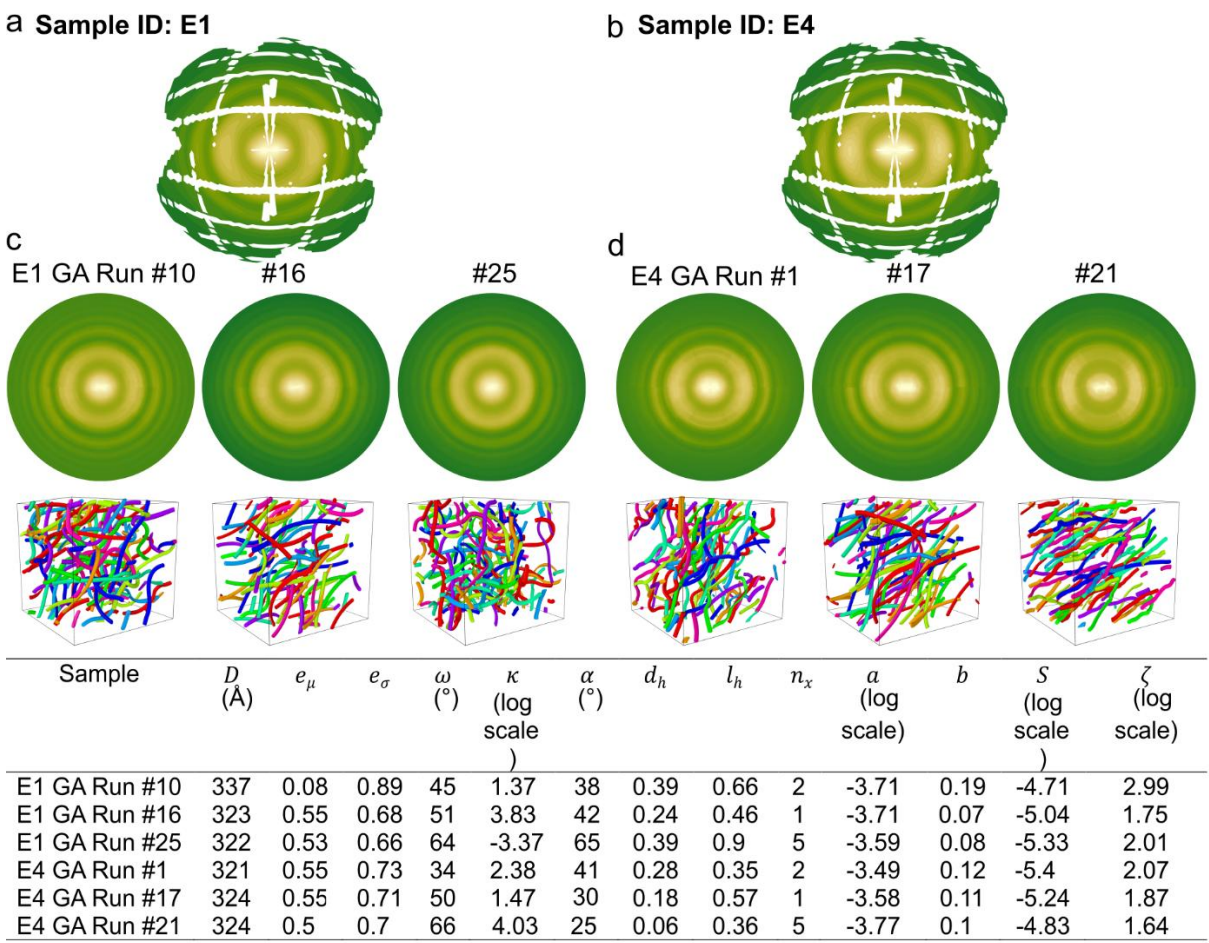

| Sample | $D$ (Å) | $e_\mu$ | $e_\sigma$ | $\omega$ (°) | $\kappa$ (log scale) | $\alpha$ (°) | $d_h$ | $l_h$ | $n_x$ | $a$ (log scale) | $b$ | $S$ (log scale) | $\zeta$ (log scale) |
|---|---|---|---|---|---|---|---|---|---|---|---|---|---|
| E1 GA Run #10 | 337 | 0.08 | 0.89 | 45 | 1.37 | 38 | 0.39 | 0.66 | 2 | -3.71 | 0.19 | -4.71 | 2.99 |
| E1 GA Run #16 | 323 | 0.55 | 0.68 | 51 | 3.83 | 42 | 0.24 | 0.46 | 1 | -3.71 | 0.07 | -5.04 | 1.75 |
| E1 GA Run #25 | 322 | 0.53 | 0.66 | 64 | -3.37 | 65 | 0.39 | 0.9 | 5 | -3.59 | 0.08 | -5.33 | 2.01 |
| E4 GA Run #1 | 321 | 0.55 | 0.73 | 34 | 2.38 | 41 | 0.28 | 0.35 | 2 | -3.49 | 0.12 | -5.4 | 2.07 |
| E4 GA Run #17 | 324 | 0.55 | 0.71 | 50 | 1.47 | 30 | 0.18 | 0.57 | 1 | -3.58 | 0.11 | -5.24 | 1.87 |
| E4 GA Run #21 | 324 | 0.5 | 0.7 | 66 | 4.03 | 25 | 0.06 | 0.36 | 5 | -3.77 | 0.1 | -4.83 | 1.64 |

**Figure 9:** *(a-b) Processed 2D SAXS profiles of E1 and E4 experimental data. (c-d) CREASE-2D output for E1 of GA Run #10, #16, #25 and E4 of GA Run #1, #17, #21 and their reconstructed 3D structures, structural features.*

*(i) Comparison of E1 and E4 samples:* Both E1 and E4 samples have (L,D)-2NapFF in de-ionized water at 10mg/ml but differ in added solvent - 10% DMSO (E1) and no added solvent (E4), respectively. **Figure 9** presents the comparison of all structural features of the best individuals for a few GA runs for E1 and E4 samples; Figure S7 presents the dendrograms of all GA runs for these two samples. For all GA runs for sample E1 and E4, we obtain similar diameter values of ~320-340 Å (Table S.3. and Table S.6); these values are similar to the diameters characterized via microscopy for (L,D)-2NapFF in de-ionized water (Figure 1). Of course,

CREASE-2D identifies other structural features that we cannot access in microscopy, e.g., shape of the tubes' cross-section (eccentricity), tubes' flexibility, tortuosity, and orientational alignment. The mean eccentricity for E1 and E4 is ~0.5 suggesting a slightly elliptical cross-section shape; the dispersity in eccentricity is also high for both E1 and E4. The tortuosity quantified by α and herding tube diameter $d_h$ is similar for E1 and E4 samples. The extent of tubes' orientational order is also similar in both samples; this similarity in E1 and E4 structures can also be seen visually in the representative structures in Figure 9. This analysis suggests that addition of 10 v/v % DMSO is not sufficient to alter the assembly of 2NapFF dipeptides seen in pure de-ionized water.

*(ii) Comparison of E2, E6, and E7 samples:* E2, E6, and E7 samples have 2NapIF in de-ionized water at 10mg/ml with added KCl at varying concentrations. The predicted data with high fitness scores from CREASE-2D for E2 and E6 (Tables S.4. and S.8, respectively) show tubes of diameters (~170-190 Å) with cross-sectional eccentricity ~0.5-0.7, suggesting an elliptical cross-section. The diameters fall in line with previous microscopy and have similar values to the analytical model fits to the azimuthally averaged 1D SAXS patterns (**SI section S.VI**), with the samples E2 and E6 showing diameters of around 140 Å. As noted in Table S.1., the samples E2 and E6 have higher KCl concentration than sample E7. At lower KCl concentration (i.e., sample E7), CREASE-2D predicts structures with higher eccentricity (values of 0.7-0.95) suggesting the formation of tapes rather than cylindrical/elliptical tubes. The diameters of E7 are also larger ~300 Å than we see for samples E2 and E6 which were in the range ~170-190 Å. The E2 and E6 samples also show higher values of orientational alignment than E7. We do not see significant differences in tortuosity among the three samples.

*(iii) Comparison of E3 and E8 samples:* These two samples have BrNapIF dipeptides in deionized water at 10mg/ml and differ only in absence or presence of NaCl salt. The CREASE-2D prediction indicates that slightly larger structures are formed by BrNapIF in the presence of a sodium salt (E8) as compared to no salt (E3). We note that the fitness values of the best individuals in the GA runs for E3 are higher than those of E8 (Tables S.5 and S.10). Sample E3 has more consistency among the values of mean cross-sectional eccentricity, orientational order parameter log κ, the 'herding' tube diameter, length, and angle α than we see for sample E8. We also make another key observation regarding the 2D SAXS profile for sample E8 (Figure 3h) which seemingly has two directions of anisotropy. However, CREASE-2D only seems to identify one direction of anisotropy (the more dominant one). This is because, during the initial phase of CREASE-2D implementation, we did not account for the possibility of multi-modal distributions of anisotropy, which is significantly more complex than a unimodal distribution. This complexity implies that the mean orientation cannot be expressed by a single parameter $\omega$, but rather by a 3D vector defined by at least two angles. We might be able to model a multi-modal distribution of orientation by combining several unimodal vMF distributions linearly. However, each mode of multi-modal distribution requires a mean direction and an orientational anisotropy parameter, which substantially increases the number of structural features. Despite these challenges, once these structural features are successfully implemented, a more extensive training and validation dataset will be necessary to develop an ML model capable of reliably predicting all the structural features within the CREASE-2D GA loop.

## V. Comparison of CREASE-2D interpretations to dipeptide nanotubes' structural understanding from other approaches

Next, we compare these CREASE-2D interpretations to interpretations we obtain from approximate analytical model fitting of the azimuthally averaged 1D SAXS profiles for selected samples and relevant work in simulations focused on assembly of dipeptides. In **S.I. section S.VI.** we present the traditional analytical model fitting and resulting quantitative interpretations.

E1 and E4 samples only differ in the added solvent amount (10% DMSO, v/v). The analytical model fits for the 1D SAXS profiles of E1 and E4 using the hollow cylinder model (Tables S11 and S12) shows no significant change in model parameters for both systems, with similar fit values within error for both samples. This suggests that the added DMSO solvent molecules at this small concentration does not alter the assembly of the dipeptides into nanotube structures in water. These experimental details and fits to 1D SAXS profiles however are unable to provide further information around flexibility, tortuosity, and orientational alignment; CREASE-2D provides this information and shows that these additional structural features - flexibility, tortuosity, and orientational alignment – are also similar for E1 and E4. Indeed, the experimental fits to the 1D SAXS data requires a hollow cylinder model and interprets the nanotube lengths to be ~170 nm. This, however, is clearly far shorter than what we observe in the cryo-TEM data; this is yet another demonstration that existing analytical models limit us to only certain shapes. As the fitting software does not have a flexible hollow cylinder available, we have interpreted the 170 nm as being the Kuhn length of the hollow cylinder as opposed to the absolute length. As such, the CREASE-2D output is able to inform much more effectively about flexibility. Further, the similar eccentricity for E1 and E4 as suggested by CREASE-2D is information that is not accessible from the fits to the 1D SAXS data. One could expect that DMSO molecules could locate themselves in the tube walls leading to plasticization; CREASE-2D data tells that this is not happening at 10% DMSO.

The data from 2NapIF samples (E2, E6, and E7) in the presence of varying equivalents of salts is in line with previous observations for samples of a similar dipeptide[28]. At lower concentrations of salt (sample E7), charge screening of the structures may lead to lateral association and give rise to the tape-like structures identified by CREASE-2D. As the concentration of salt increases, new self-assembled structures are formed (samples E2 and E6). In contrast to 2NapIF, we see smaller differences for BrNapIF in presence of salt (samples E3 and E8); this highlights how the effect of adding salt also depends on molecular structure of the dipeptide. Higher radii of nanotubes are observed when salt is present, which could result from charge screening-induced association of the tubes.

To further demonstrate the power of CREASE-2D's interpretations of the entire SAXS profiles, we discuss briefly other molecular simulation work focused on related dipeptide systems to show the limitations of molecular simulations in providing the distributions of structural features that CREASE-2D has provided us. For example, simulations of the structures formed by a naphthalene-dipeptide have been carried out [29], but these studies have predicted an incorrect morphology (worm-like micelles) as opposed to the nanotubes formed [30]. Had these simulations been compared to the available SAXS data, this inconsistency between simulations and experiments would have become immediately apparent.

While CREASE-2D interpretations of structural features lacks a physical reasoning (e.g., the molecular driving forces) for the observed structures, molecular simulations can describe the types of driving forces and molecular packing *within* the aggregates formed by the functionalized dipeptides and experimental data can be correlated with the outcomes. However, again, simulations provide information at smaller-lengths (i.e., molecular-level information) as opposed to information on the ensemble of aggregates (leading to distributions of structural features) [31]. Examples exist where simulations are used to understand the morphology that is likely adopted, for example vesicles and nanotubes being formed by diphenylalanine [32-34]. One could explain the experimentally observed structures with such simulations, but the high computational intensity of simulations limits them to only single aggregates. In all these cases, limited understanding of experimental systems arises in terms of explaining larger-scale (i.e. bulk) systems. CREASE-2D approach allows us to understand data from concentrated samples and provides valuable information in terms of distributions of observed structural shapes, sizes, and alignment. CREASE-2D also directly uses experimental data as input unlike most other molecular simulation work in this field.

Overall, the structural interpretation of the samples obtained from CREASE-2D analysis of 2D SAXS profiles provide an unprecedented level of understanding on the dipeptide self-assembly which has not been possible through traditional analytical model fits to the 1D SAXS data or simulations. For instance, discrimination between more tape-like structures or cylindrical structures is hard to ascertain from analysis of the 1D and 2D SAXS without the supporting use of microscopy techniques; however, as noted before, microscopy techniques do not provide conclusive information on the tubes' cross-section, orientational alignment, and tortuosity of the assembled dipeptide tubes.

## VI. Conclusions

In this work, we have demonstrated the extension and application of the CREASE-2D method to analyze the experimentally obtained 2D scattering profiles of dipeptide micellar solutions. CREASE-2D interprets the profiles in the form of distributions of structural features corresponding to sizes and cross-sectional shapes of the tubular micelles, the stiffness and tortuosity of the tubes, and presence/absence of orientational alignment. These quantitative structural features are then used to reconstruct the representative 3D structures of these dipeptide micellar systems. Having CREASE-2D method for analysis of the entire 2D SAXS data in place of (or in addition to) traditional analytical model fitting of azimuthally averaged 1D SAXS data is highly enabling for the field. In traditional analytical model fits of 1D SAXS data it can be difficult to be certain which models are most appropriate and what fits would be correct; CREASE-2D overcomes such challenges and provides all possible structural interpretations to understand the data. In particular, for structures with orientational alignment either driven by thermodynamics or due to processing, analysis of the azimuthally averaged SAXS data can be too approximate or incorrect as that assumes isotropy in the structure. Having the ability to analyze all of the 2D SAXS data without such averaging is powerful. For both isotropic and anisotropic structures, CREASE-2D is able to identify distributions of relevant structural features including ones that cannot be identified with existing analytical models (e.g., assembled tube cross-sectional eccentricity, orientational order), which is a significant and a transformative step forward for the soft materials community.

## Acknowledgements

We are grateful for the financial support from Department of Energy BES SC0023264. We also acknowledge the use of DARWIN computing system for some of the computing in this work: DARWIN—A Resource for Computational and Data-Intensive Research at the University of Delaware and in the Delaware Region, which is supported by the NSF under grant no. 1919839, Rudolf Eigenmann, Benjamin E. Bagozzi, Arthi Jayaraman, William Totten, and Cathy H. Wu, University of Delaware, 2021. Simona Bianco thanks the University of Glasgow for funding. Dave Adams thanks the Leverhulme Trust (RPG-2022-324) for funding.

## Data Availability

The open-source code and experimental data used for CREASE-2D Analysis of SAXS data for Supramolecular dipeptide micelles is available on https://github.com/arthijayaraman-lab/CREASE-2D-Tubes-Raw-SAXS-Analysis .